\documentstyle[12pt]{article}
\title{Higgs' mass around 115 GeV would testify for composite Higgs} 
\author{B.A. Arbuzov\\
{\it Skobeltsyn Institute for Nuclear Physics, Moscow State 
University}\\ {\it 119899 Moscow, Russia}}
\date{}
\newcommand{\be}{\begin{equation}}
\newcommand{\ee}{\end{equation}}
\newcommand{\beq}{\begin{eqnarray}}
\newcommand{\eeq}{\end{eqnarray}}
\newcommand{\nn}{\nonumber}
\newcommand{\bi}{\bibitem}
\begin{document}
\maketitle
\begin{center}
Higgs' mass 115 GeV testifies for composite Higgs
\end{center}
\begin{quote}
Theory with $SU(2)\times U(1)$ gauge invariant electroweak Lagrangian 
describing standard interaction of massless quark doublet without 
elementary scalar Higgs sector is considered.  
We show in the main order of $1/N_c$ expansion, that there exists a 
solution, 
which breaks initial symmetry and is accompanied by 
an appearance of zero mass composite scalar doublet. 
The electroweak loop corrections 
shift its mass to a tachyon value and thus the scalar serves as 
a substitute for usual elementary Higgs. The $t$-quark mass also 
arise. The problem of adjusting its observable mass leads to prediction 
of triple gauge constant $\lambda_V\,\simeq\,-0.034$ without 
contradiction with experimental data. The mass of the surviving 
composite Higgs scalar is predicted to be  
$M_H = 117 \pm 7\,GeV$, that agrees with recent 
indications of $M_H \simeq 115\,GeV$. The resulting  
theory contains realistic $W,\,Z,\,t$ masses, composite massive Higgs 
and massless other particles. Parameters used 
are cut-off $\Lambda\,\simeq\,3500\,GeV$, weak gauge constant $g$ 
and mixing angle $\theta_W$.

\end{quote}

\section{Introduction}

The widely popular Higgs mechanism~\cite{Higgs} of the electroweak symmetry 
breaking 
needs initial scalar fields, which look rather less attractive, than 
the well-known gauge interactions of vector and spinor fields. 
Experimental facilities now approach the region of possible discovery of 
the Higgs. There are even serious indications in favour of the Higgs mass 
$M_H\,=\,115^{+1.3}_{-0.9}\,GeV$~\cite{New}. 
In view of these considerations it may be useful to 
study once more possibilities, which differ from the standard Higgs 
mechanism and to look for experimental consequences of these possibilities.

In the present work\footnote[1]{The work is supported in part by 
grant "Universities of Russia" 990588.}
we consider a model, which might serve a substitute for the Standard
Model in respect to symmetry breaking. We are basing here on the proposal, 
which was done in work~\cite{Arb00}, the main point of which consists in 
application of N.N. Bogolyubov method of quasi-averages~\cite{Bog} 
to a simple model. Let us summarize the essence of the 
approach~\cite{Arb00} with some necessary corrections. 
Namely we consider $U(1)$ 
massless gauge field $A_\mu$ and also massless spinor field $\psi$, 
which interact in the following way
\beq
& &L\,=\,\frac{\imath}{2}\bigl(\bar \psi \gamma_\rho \partial_\rho \psi\,- 
\,\partial_\rho \bar \psi \gamma_\rho \psi \bigr)\,-\,\frac{1}{4}
A_{\mu \nu}A_{\mu \nu}\,+\nn\\
& &+\,e_{L}\bar \psi_L \gamma_\rho \psi_L A_\rho\,
+\,\,e_{R}\bar \psi_R \gamma_\rho \psi_R A_\rho\,;\label{init}\\
& &A_{\mu\nu}\,=\,\partial_\mu A_\nu\,-\,\partial_\nu A_\mu\,;\nn
\eeq
where as usually
$$
\psi_L\,=\,\frac{1+\gamma_5}{2}\,\psi\,;\qquad 
\psi_R\,=\,\frac{1-\gamma_5}{2}\,\psi\,.
$$
We will not discuss here the problem of 
the triangle axial anomaly, bearing in mind that in a 
more realistic model one always can arrange cancellation of the 
anomalies by a suitable choice of fermions' charges, in the same way 
as it occurs in the Standard Model.

Now we start to apply Bogolyubov quasi-averages method~\cite{Bog}
\footnote[2]{The quasi-averages method was first applied to 
quantum field theory problems in work~\cite{ATF}.}. In view 
of looking for symmetry breaking we add to~(\ref{init}) additional term
\be
\epsilon \cdot \bar \psi_L \psi_R\,\bar \psi_R \psi_L\,.\label{eRL}
\ee

Now let us consider the theory with $\epsilon \neq 0$, calculate necessary 
quantities (averages) and only at this stage take limit $\epsilon \to 0$. 
In this limit, according~\cite{Bog}, we come to quasi-averages, which 
not always coincide with the corresponding averages, which one obtains 
directly from the initial Lagrangian~(\ref{init}).

Because of additional term~(\ref{eRL}) the following additional effective 
terms in Lagrangian~(\ref{init}) inevitably appear
\beq
& &\Delta L_x\,=\,x\cdot \bar \psi_L \gamma_\rho \psi_L\,\bar \psi_R \gamma_
\rho \psi_R\,; \qquad \Delta L_y\,=\,\frac{y}{2}\cdot \bar \psi_L \gamma_
\rho \psi_L\,\bar \psi_L \gamma_\rho \psi_L\,;\nn\\
& &\Delta L_z\,=\,\frac{z}{2}\cdot \bar \psi_R \gamma_\rho \psi_R\,
\bar \psi_R \gamma_\rho \psi_R\,.
\label{psi}
\eeq
The corresponding vertices should have form-factors, which define 
effective cut-off $\Lambda$. 
The origin of the cut-off is connected with self-consistent 
solution of the corresponding dynamical equations. A realization of 
this mechanism will be shown below in Appendix A. 
In the present work we just set upper limit at Euclidean 
$p^2\,=\,\Lambda^2$ in momentum integrals.  

We consider compensation equations~\cite{Bog, Bog2} 
(in other words, gap equations) for $x,\,y,\,z$ in one-loop 
approximation (see Fig. 1), neglecting for the moment gauge field 
$A_\mu$ exchange loops, and obtain following set of equations
\beq
& & x\,=\,-\,\frac{\epsilon}{2}\,
+\,\frac{\Lambda^2}{16 \pi^2}\Bigl(\,-\, 3\,x^2\,-\,2\,x\,(y + z)
\Bigr)\,;\nn\\
& & y\,=\,\frac{\Lambda^2}{16 \pi^2}\Bigl( -\,x^2\,\Bigr)\,;\qquad 
z\,=\,
\frac{\Lambda^2}{16 \pi^2}\Bigl( -\,x^2\,\Bigr)\,;\label{set}
\eeq 
Let $X,\,Y,\,Z$ be dimensionless variables 
\be
X\,=\,x\,\frac{\Lambda^2}{16 \pi^2}\,;\qquad
Y\,=\,y\,\frac{\Lambda^2}{16 \pi^2}\,;\qquad
Z\,=\,z\,\frac{\Lambda^2}{16 \pi^2}\,;\label{X}
\ee
Let us look for  solutions of set~(\ref{set}). At this stage we also 
set $\epsilon \to 0$. There is, of course, trivial solution
$x\,=\,y\,=\,z\,=\,0\,$. 
In addition we have two nontrivial solutions. 
\beq
& &Z_1\,=Y_1\,=\,-\,1\,;\qquad X_1\,=\,1\,;\label{1} \\ 
& &Z_2\,=Y_2\,=\,-\,\frac{1}{16}\,;\qquad X_2\,=\,-\,\frac{1}{4}\,.\label{2}
\eeq

As we shall see further, just the second solution~(\ref{2}) 
will be the most interesting.  

Now let us consider scalar bound states $(\bar \psi_L \psi_R,\,\bar \psi_R
\psi_L)$. Without $e^2$ corrections we have from Bethe-Salpeter 
equation in one-loop approximation (see Fig. 2)
\beq
& &G\,=\,-\,4\,X\,F(\xi)\,G\,;\qquad \xi\,=\,\frac{k^2}{4 \Lambda^2}\,;
\qquad\mu\,=\,\frac{m^2}{\Lambda^2};\label{eq0}\\
& &F(\xi)\,=\,1\, -\, \frac{5}{2}\,\xi\, +\, 2\xi\,
\log\frac{4 \xi}{1 + \xi}\,;\qquad \mu \ll \xi\,<\,1\,;\nn\\
& &F(\xi)\,=\,1\, +\, \frac{1}{3}\,\xi\, +\, 2\xi\,
\log \mu\,+\,O(\xi\mu,\,\xi^2)\,;\qquad \xi\leq \mu \,.\nn
\eeq
where $G\,=\,const$ is just the Bethe-Salpeter wave function. Here 
$m$ means a fermion mass, which will be shown to be nonzero,
and $k^2$ is the scalar state Euclidean momentum squared, that is 
$k^2 > 0$ means tachyon mass of the scalar. 
Function $F(\xi)$ decreases 
from the value $F(0) = 1$ with $\xi$ increasing. We see, that 
for solution~(\ref{2}) we have bound state with $k^2 = 0$ in full 
correspondence with Bogolyubov-Goldstone theorem~\cite{Bog2}, 
\cite{Gold}. As for the first solution~(\ref{1}), there is no 
solution of Eq.~(\ref{eq0}) at all.  So we have to  concentrate 
our attention on the solution~(\ref{2}). Note, that there is 
an additional argument in favour of solution~(\ref{2}). Namely, 
values $X$ and especially $Y,\,Z$ are small enough, so we may 
expect, that many-loop terms will not influence results strongly. 

Now let us take into account vector boson corrections. Equation 
for the bound state~(\ref{eq0}) is modified due to two sources. 
The first one corresponds to vector boson exchange loops in 
set~(\ref{set}). Here corrections appear, for example, of such form
$$
\frac{6\,e_L\,e_R\,x}{16\,\pi^2}\,\log\frac{\Lambda^2}{p^2}\,.
$$
Here $p$ is the momentum of integration in 
Eq.~(\ref{eq0}). In the present note we will restrict ourselves by 
main logarithmic approximation, in which such terms do not contribute 
to results. Indeed, due to simple relation
\be
\int_0^{\Lambda^2}\,x^n\,\log\frac{\Lambda^2}{x}\,dx\,=\,\frac{
(\Lambda^2)^{n+1}}{(n+1)^2}\,;\nn
\ee
there is no logarithms in final expressions. So the only important 
contribution consists in loop $e^2$ 
corrections to Eq.~(\ref{eq0}). In Landau gauge there 
is only one such contribution: the triangle 
diagram. We draw at Fig. 2 only this diagram, but the 
gauge invariance of the result is checked by direct calculations in 
an arbitrary gauge.  So we now have
\be
G\,=\,F(\xi)\,G\,+\,
\frac{3\, e_L\, e_R}
{16\, \pi^2}\,\log \frac{\Lambda^2}{m^2}\,G\,.\label{eqe}
\ee
We see, that for small $e_L,\,e_R$ possible eigenvalues $\xi$ are 
also small. Then we have following condition for an eigenvalue
\be
F(\xi)\,+\,\frac{3\, e_L\, e_R}{16\, \pi^2}\,\log \frac{\Lambda^2}{m^2}\,
=\,1\,;\label{eigen}
\ee
We see, that there is tachyon bound state in case $e^2$  
contribution being positive. 
Really, the eigenvalue condition for small $e_i^2$ reads
\be
k^2\,=\,m_0^2\,=\,\frac{3\, e_L\, e_R}{8\, \pi^2}\,\Lambda^2\,
.\label{m0}
\ee
The result corresponds to main logarithmic approximation. 
Thus provided $(e_L\, e_R) > 0$ we have scalar complex tachyon 
$\phi$ with negative mass squared $-\,m_0^2$.
 
We have the following vertices of interaction of $\phi$ with spinors
\be
G\,\Bigl(\bar \psi_R\,\psi_L\,\phi\,+
\,\bar \psi_L\,\psi_R\,\phi^*\Bigr)\,.\label{fint}
\ee
Normalization condition of Bethe-Salpeter equation gives
\be
G^2\,=\,\frac{16\,\pi^2}{\log\,(\Lambda^2/m^2)}\,.\label{c}
\ee

Then we calculate box diagram with four scalar legs. This gives us 
effective constant $\lambda$, which enters into additional term 
\be
\Delta\,L\,=\,-\,\lambda\,(\phi^*\,\phi)\,(\phi^*\,\phi)\,;
\qquad
\lambda\,=\,\frac{G^4}{16\,\pi^2}\,\log\,(\Lambda^2/m^2).\label{lam}
\ee
Now we come to the usual Higgs model~\cite{Higgs} with $m_0^2$~(\ref{m0}), 
$\lambda$~(\ref{lam}) and $\phi$ charge $e_L\,-\,e_R$.

Thus from expressions~(\ref{m0}, \ref{lam}) we have usual 
vacuum average of $\sqrt{2}\,Re\,\phi\,=\,\eta$
\be
\eta^2\,=\,\frac{m_0^2}{\lambda}\,=\,\frac{3\,e_L\,e_R}{128\,\pi^4}\,
\Lambda^2\,\log\,\frac{\Lambda^2}{m^2}\,.\label{eta}
\ee

The vector boson mass duly arises and it reads as follows
\be
M^2\,=\,\frac{3\,(e_L - e_R)^2\,e_L\,e_R}{128\,\pi^4}\,\Lambda^2\,
\log\,\frac{\Lambda^2}{m^2}\,.\label{M}
\ee

Interaction~(\ref{fint}) leads to spinor mass $m$
\be
m\,=\,\frac{G\,\eta}{\sqrt{2}}\,;\quad
m^2\,=\,\frac{3\,e_L\,e_R}{8\,\pi^2}\,\Lambda^2\,.\label{m}
\ee

Thus, we obtain the result, that initially massless model of interaction 
of a spinor with a vector becomes after the symmetry breaking just a 
close analog of the Higgs model. We have now vector boson mass~(\ref{M}), 
spinor mass ~(\ref{m}) and a scalar bound state with mass $\sqrt{2}\,m_0$, 
\be
m^2_H\,=\,2\,m_0^2\,=\,\frac{3\, e_L\, e_R}{4\, \pi^2}\,\Lambda^2\,.
\label{mH}
\ee

We would formulate qualitative result of the simple model study as follows: 
in the massless model with Lagrangian~(\ref{init}) with 
$(e_L \cdot e_R)\,>\,0$ there arises fermion-antifermion condensate, which 
defines masses $M,\,m,\,M_H$ according to~(\ref{M}, \ref{m}, \ref{mH}).

Note, that variants of dynamical breaking of the electroweak symmetry 
without elementary scalars were considered in various aspects 
(see, e.g. paper~\cite{Arb92}). The possibility of  
scalars being composed of fundamental spinors was considered e.g. in 
well-known paper~\cite{Ter}.

\section{Realistic model in the leading order of $1/N_c$ expansion}

Here we 
will proceed in the same line, but choose more realistic model, including 
one coloured left doublet $\psi_L\,=\,(t_L,\,b_L)$ and two right singlets 
$t_R,\,b_R$, which simulate the heaviest quark pair. They interact 
with $SU(2)\times U(1)$ gauge bosons in standard way, so that the initial 
Lagrangian corresponds to the Standard Model with one heavy quark 
(initially massless) generation without Higgs sector and looks like

\beq
& &L\,=\,
\frac{\imath}{2}\bigl(\bar \psi_L \gamma_\rho \partial_\rho \psi_L\,- 
\,\partial_\rho \bar \psi_L \gamma_\rho \psi_L \bigr)\,+\,
\frac{\imath}{2}\bigl(\bar t_R \gamma_\rho \partial_\rho t_R\,- 
\,\partial_\rho \bar t_R \gamma_\rho t_R \bigr)\,+\nn\\
& &+\,\frac{\imath}{2}\bigl(\bar b_R \gamma_\rho \partial_\rho b_R\,- 
\,\partial_\rho \bar b_R \gamma_\rho b_R \bigr)\,-
\,\frac{1}{4}W_{\mu \nu}^a W_{\mu \nu}^a\,
-\,\frac{1}{4}B_{\mu \nu} B_{\mu \nu}\,-
\,\frac{g}{2}\,\bar \psi_L \tau^a\gamma_\rho \psi_L W^a_\rho\,
-\nn\\
& &-\,\frac{g\tan \theta_W}{6 }\bar \psi_L \gamma_\rho \psi_L B_\rho\,
-\,\frac{2 g \tan \theta_W}{3}\bar t_R \gamma_\rho t_R B_\rho\,+\,
\frac{g \tan \theta_W}{3}\bar b_R \gamma_\rho b_R B_\rho\,;\label{initr}\\
& &W_{\mu\nu}^a = \partial_\mu W_\nu^a - \partial_\nu W_\mu^a + 
g\,\epsilon^{abc}\,W^b_\mu\,W^c_\nu\,;\;
B_{\mu\nu} = \partial_\mu B_\nu - \partial_\nu B_\mu\,;\;g = 
\frac{e}{\sin \theta_W}.\nn
\eeq
Notations here are usual ones and quarks $t$ and $b$ 
are colour triplets.  

Let us look for spontaneous four-fermion interactions 
\beq
& &x_1\,\bar \psi_L^\alpha \gamma_\rho \psi_{L \alpha}\,
\bar t_R^\beta \gamma_\rho t_{R \beta};\quad
x_2\,\bar \psi_L^\alpha \gamma_\rho \psi_{L \alpha}\,
\bar b_R^\beta \gamma_\rho b_{R \beta};\nn\\
& &\bar x_1\,\bar \psi_L^\alpha \gamma_\rho \psi_{L \beta}\,
\bar t_R^\beta \gamma_\rho t_{R \alpha};\quad
\bar x_2\,\bar \psi_L^\alpha \gamma_\rho \psi_{L \beta}\,
\bar b_R^\beta \gamma_\rho b_{R \alpha};\nn\\
& &\frac{y_1}{2}\,\bar \psi_L^{\alpha\,s} \gamma_\rho \psi_{L \alpha\,s}\,
\bar \psi_L^{\beta\,r}\gamma_\rho \psi_{L \beta\,r}\,;\quad
\frac{y_2}{2}\,\bar \psi_L^{\alpha\,s} \gamma_\rho \psi_{L \alpha\,r}\,
\bar \psi_L^{\beta\,r} \gamma_\rho \psi_{L \beta\,s}\,;\label{nset}\\
& &\frac{z_1}{2}\,\bar t_R^\beta \gamma_\rho t_{R \beta}\,
\bar t_R^\alpha \gamma_\rho t_{R \alpha}\,;\quad
\frac{z_2}{2}\,\bar b_R^\alpha \gamma_\rho b_{R \alpha}\,
\bar b_R^\beta \gamma_\rho b_{R \beta}\,;\nn\\
& &z_{12}\,\bar t_R^\beta \gamma_\rho t_{R \beta}\,
\bar b_R^\alpha \gamma_\rho b_{R \alpha}\,;\quad
\bar z_{12}\,\bar t_R^\beta \gamma_\rho t_{R \alpha}\,
\bar b_R^\alpha \gamma_\rho b_{R \beta}\,;\nn\\
& &\bar y_1\,=\,y_2\,;\quad \bar y_2\,=\,y_1\,;\quad
\bar z_i\,=\,z_i\,.\nn
\eeq
Here $\alpha,\,\beta$ are colour indices and $s,\,r$ are weak isotopic 
indices. Symbol $\bar a$ means interchange of colour summation just in 
the way as it is done for $x_1,\,x_2$. In the same way, as earlier
~(\ref{X}) we introduce dimensionless variables
\be
X_i\,=\,\frac{x_i\, \Lambda^2}{16\,\pi^2}\,;\quad Y_i\,=\,
\frac{y_i\, \Lambda^2}{16\,\pi^2}\,;\quad
Z_i\,=\,\frac{z_i\, \Lambda^2}{16\,\pi^2}\,.
\label{Xi}
\ee
From diagram set of equations Fig. 1 we could obtain a complicated 
set of algebraic equations. Its form is marvelously simplified 
in the main 
order of $1/N_c$ expansion. So here we consider just this approximation. 
Thus for the beginning 
we get equations for $Y_1,\,Y_2$
\beq
& &Y_1\,=\,\frac{N_c}{2}\,\Bigl(-\,Y_2^2-2 Y^2_1-X^2_1-X^2_2-2 Y_1\,Y_2
\Bigr)\,;\nn\\
& &Y_2\,=\,\frac{N_c}{2}\,\Bigl(-\,Y_2^2-2 Y^2_1-X^2_1-X^2_2-2 Y_1\,Y_2
\Bigr)\,.\label{y}
\eeq
This means, that $Y_1\,=\,Y_2\,=\,Y$. For other variables we have
\beq
& &X_1\,=\,-\,N_c\,\Bigl(3 X_1\,Y+2 X_1 Z_1 + X_2 Z_{12}\Bigr)\,;\nn\\
& &X_2\,=\,-\,N_c\,\Bigl(3 X_2\,Y+2 X_2 Z_2 + X_1 Z_{12}\Bigr)\,;\nn\\
& &Y\,=\,-\,\frac{N_c}{2}\,\Bigl(5 Y^2+X_1^2+X_2^2\Bigr)\,;\quad
\bar Y\,=\,Y;\;\bar Z_1\,=\,Z_1;\;\bar Z_2\,=\,Z_2;\label{Nc}\\
& & Z_1\,=\,-\,\frac{N_c}{2}\,\Bigl(4\,Z_1^2 + 2\,X_1^2 + Z^2_{12}  
\Bigr);\quad Z_2\,=\,-\,\frac{N_c}{2}\,\Bigl(4\,Z_2^2 + 2\,X_2^2 + 
Z^2_{12}\Bigr);\nn\\
& & Z_{12}\,=\,-\,2\,N_c \Bigl( Z_1\,Z_2 + Z_1\,Z_{12} + Z_2\,Z_{12}
\Bigr);\quad \bar Z_{12}\,=\,-\,N_c\,\bar Z_{12}^2\,.\nn\\
& &\bar X_1\,=\,-4\,N_c\,\bar X_1^2\,;\quad 
\bar X_2\,=\,-4\,N_c\,\bar X_2^2\,;\label{nt}
 \eeq
The most important are equations for $\bar X_{1,2}$, which we mark  
separately. If we consider Bethe-Salpeter equations for 
$\bar \psi_L t_R$ and $\bar \psi_L b_R$ scalar states again 
in the main order of $1/N_c$ expansion, we have 
respectively
\be
G_1\,=\,-\,4\,N_c\,\bar X_1\,F(\xi)\,G_1\,;\quad
G_2\,=\,-\,4\,N_c\,\bar X_2\,F(\xi)\,G_2\,;\qquad 
\xi\,=\,\frac{k^2}{4 \Lambda^2}\,.\label{eqn}
\ee
Here $F(\xi)$ is the same function as earlier~(\ref{eq0}).
This means that in case of a non-trivial solution of one of equations
~(\ref{nt}) we  have zero mass Bogolyubov-Goldstone scalar doublet. 
Interaction with gauge vector  fields according to~(\ref{initr}) shifts 
the level to positive or negative values of the mass squared. 
Because of the bound state consisting of left and right spinors, 
only interactions of $B$-boson enter to diagram Fig. 2. 
So if $\bar X_1\,=\,-\,1/4 N_c$, we  have tachyon state with
\be
m_0^2\,=\,\frac{g^2 \tan^2\theta_W}{24\,\pi^2}\,\Lambda^2\,.\label{m0n}
\ee
For other possibility, i.e. $\bar X_2\,=\,-\,1/4 N_c$ ($\bar \psi_L\, b_R$ bound 
state), we have normal sign mass squared
$$
m_{L b}^2\,=\,\frac{g^2 \tan^2\theta_W}{48\,\pi^2}\,\Lambda^2\,.
$$
As we already know, the first possibility leads to Higgs-like symmetry 
breaking, that is to negative minimum of the effective potential. As 
for the second case, there is no minima, so here we should take 
trivial solution $\bar X_2\,=\,0$. It is very important point. Really, 
we here obtain the explanation of why the $t$-quark is heavy and the 
$b$-quark is light. Only the first possibility corresponds to 
tachyon and so gives condensate, leading to creation of masses including 
$t$-quark mass. The fact is connected with signs of interaction terms in 
~(\ref{initr}). For us here only $\bar X_1$ is important. It is easily 
seen from set~(\ref{Nc}) that for all other variables we can take 
trivial solution $X_1 = X_2 = ... = 0$. One should expect, that the 
situation might change in the next approximations. 

Next step is normalization of the bound state. Again following previous 
considerations~(\ref{fint}, \ref{c}) we have the following normalization condition 
\be
\frac{N_c\,G^2_1}{16\,\pi^2}\,\log\frac{\Lambda^2}{m^2}\,=\,1\,;
\label{norm}
\ee
Now we obtain fourfold scalar interaction constant $\lambda$
\be
\lambda\,=\,\frac{N_c\,G^4_1}{16\,\pi^2}\,\log\frac{\Lambda^2}{m^2}\,.
\label{lamn}
\ee

For the moment we have everything for effective Higgs mechanism in 
our variant of the Standard Model. We immediately come to classical 
scalar field density $\eta$
\be
\eta^2\,=\,\frac{m_0^2}{\lambda}\,;\label{etan}
\ee
where parameters are uniquely defined~(\ref{m0n}, \ref{norm}, \ref{lamn}). 
We also easily see, that constants of gauge interactions of the 
scalar doublet are just $g$ for interaction with $W$ and 
$g'\,=\,g\,\tan \theta_W$ for interaction with $B$ as usually. 
Now we have at once $W,\,Z$ and $t,\,b$-quark masses
\be
M_W\,=\,\frac{g\,\eta}{2}\,;\quad M_Z\,=\,\frac{M_W}{\cos\theta_W}\,;
\quad m_t\,=\,\frac{G_1\,\eta}{\sqrt{2}}\,;
\quad m_b\,=\,0\,.\label{MWt}
\ee
Mass of surviving Higgs scalar 
\be
M_H^2\,=\,2\,m_0^2\,=\,2\,\lambda\,\eta^2\,=\,
\frac{N_c\,G_1^2\,m_t^2}{4\,\pi^2}\,\log\frac{\Lambda^2}{m_t^2}\,.
\label{Higgsn}
\ee
From~(\ref{m0n}) we have
$$
\Lambda^2\,=\,\frac{3\,\pi \cos^2\theta_W}{\alpha}\,M_H^2\,;
$$
where as usually $\alpha$ is the fine structure constant. 
From here and from~(\ref{Higgsn}) we have 
\be
R\,=\,\frac{3\,G_1^2}{4\,\pi^2}\,\biggl(\log R\,+\,\log\frac{3\,\pi\,
\cos^2\theta_W}{\alpha}\biggr)\,;\quad R\,=\,\frac{M_H^2}{m_t^2}\,.
\label{!}
\ee

Now let us discuss phenomenological aspects. We know all masses but 
that of the Higgs particle. To obtain it let us proceed as follows. 
From~(\ref{MWt}) we have 
\be
G_1\,=\,\frac{g\,m_t}{\sqrt{2}\,M_W}\,;\quad g\,=\,\sqrt{\frac
{4\,\pi\,\alpha}{\sin^2\theta_W}}\,=\,0.648;\quad \alpha(M_Z)\,=\,\frac{1}{129}\,.
\label{G11}
\ee
Then for values $m_t\,=\,174 \pm 5\,GeV$, $M_W\,=\,80.3\,GeV$ we have 
$G_1\,=\,0.993 \pm 0.032$. Substituting this into~(\ref{!}) we have from 
solution of the equation 
\be
M_H\,=\,117.1\, \pm\, 7.4\,GeV\,.\label{hor}
\ee
For example, for $m_t\,=\,173\,GeV$ we obtain $M_H\,=\,115.6\,GeV$.
Result~(\ref{hor}) support indications of $115\,GeV$ Higgs~\cite{New}. 

However value $G_1\,\simeq\, 1$ does not fit normalization condition~
(\ref{norm}). From~(\ref{norm}) $G_1^2$ is rather about $8$ than 
about $1$. So we conclude, that condition~(\ref{norm}) contradicts 
to the observed $t$-quark mass. We would emphasize , that the 
normalization condition is the only flexible point in our approach. 
The next section will be devoted to discussion of this problem.

\section{Normalization condition}

Scalar bound state in our approach consists of anti-doublet $(t_L,\,b_L)$ 
and singlet $t_R$. It evidently has weak isotopic spin $1/2$. Let us denote 
its field as $\phi$
\be 
\phi\,=\,\Bigl( 
\bar b_L\,t_R \quad\bar t_L\,t_R \Bigr) \,;\qquad \phi^+\,=\,
\Bigl(  \bar t_R\, b_L\quad\bar t_R\, t_L\Bigr)\,.\label{f}
\ee
From  the previous results we have the following terms 
in the effective Lagrangian
\beq
& &N\,\frac{\partial \phi^+}{\partial x^\mu}\, 
\frac{\partial \phi}{\partial x^\mu}\,+\, m_0^2\,\phi^+ \phi \,-
\,\lambda\,(\phi^+ \phi)^2 \,;\label{N}\\
& &N\,=\,\frac{N_c\,G^2_1}{16\,\pi^2}\,\log\frac{\Lambda^2}{m^2}\,;\nn
\eeq
where parameters are defined in~(\ref{m0n}, \ref{lamn}). As for 
normalization parameter $N$, it does not satisfy us. 
So we may study a possibility of quasi-averages effect 
for $\phi$ interactions, which can change 
a coefficient afore the kinetic term. So, let us consider the following 
possible interactions in addition to~(\ref{N})
\beq
& &L_{int}(W)\,=\,\frac{\xi_{01}}{2}\,\frac{\partial \phi^+}{\partial x^\mu}\, 
\frac{\partial \phi}{\partial x^\nu}\,W^a_{\mu \rho}\,W^a_{\nu \rho}\,+\,
\frac{\eta_{01}}{2}\,\frac{\partial \phi^+}{\partial x^\mu}\, 
\frac{\partial \phi}{\partial x^\mu}\,W^a_{\nu \rho}\,W^a_{\nu \rho}\,;
\label{ffWW}\\
& &L_{int}(B)\,=\,\frac{\xi_{00}}{2}\,\frac{\partial \phi^+}{\partial x^\mu}\, 
\frac{\partial \phi}{\partial x^\nu}\,B^a_{\mu \rho}\,B^a_{\nu \rho}\,+\,
\frac{\eta_{00}}{2}\,\frac{\partial \phi^+}{\partial x^\mu}\, 
\frac{\partial \phi}{\partial x^\mu}\,B^a_{\nu \rho}\,B^a_{\nu \rho}\,.
\label{ffBB}
\eeq
In obtaining compensation equations we take into account possibility 
of triple $W$ gauge bosons coupling, which was discussed in various 
papers (e.g.~\cite{Hag}) and, in particular, in a variant of dynamical breaking 
of the electroweak symmetry, which the author considered some time ago~
\cite{Arb92}. In this variant the Bogolyubov quasi-averages method is 
applied to possible origin of the triple interaction
\be
\frac{g \lambda_V}{M_W^2}\,\epsilon_{a b c}\,W_{\mu \nu}^a\,W_{\nu \rho}^b\,
W_{\rho \mu}^c\,;\label{WWW}
\ee
and corresponding one-loop compensation equation looks like (see Fig. 3)  
\be
\lambda_V\,=\,\lambda_V\,\Biggl(\frac{g \lambda_V\,}{M_W^2}\Biggr)^2\,
\frac{\Lambda^4}{128\,\pi^2}\,;\label{compW}
\ee
We have solutions of this equation: trivial one $\lambda_V\,=\,0$ 
and two non-trivial ones 
\be
\lambda_V\,=\,\pm\,\lambda_0\,;\quad 
\lambda_0\,=\,\frac{8\,\sqrt{2}\, M_W^2}{g\,\Lambda^2}\,.\label{lV}
\ee
In the following we assume, that genuine value of $\lambda_V$ may 
differ from value~(\ref{lV}) in the range of 15-20\%.

Set of equations in one-loop approximation according to diagrams 
presented at Fig. 4 looks like
\beq
& &\xi_1\,=\,-\,\frac{1}{3}\,\xi_1^2\,-\, \frac{1}{3}\,\xi_1\,\eta_1\,-
\, \frac{1}{12}\,\eta_1^2\,
+\,\frac{4}{3}\, \zeta\,\xi_1 \,;\nn\\
& &\eta_1\,=\,-\,\frac{1}{48}\,\xi_1^2\,-\, \frac{1}{12}\,\xi_1\,\eta_1\,-\,
\frac{1}{12}\,\eta_1^2\,+\,4\,\zeta\,\biggl( \frac{5}{6}\,\xi_1\,+\,2\,
\eta_1 \biggr)\,;\label{xi1}\\
& &\xi_0\,=\,-\,\frac{1}{3}\,\xi_0^2\,-\, \frac{1}{3}\,\xi_0\,\eta_0\,-
\, \frac{1}{12}\,\eta_0^2\,;\label{xi0}\\
& &\eta_0\,=\,-\,\frac{1}{48}\,\xi_0^2\,-\, \frac{1}{12}\,\xi_0\,\eta_0\,-\,
\frac{1}{12}\,\eta_0^2\,;\nn\\
& &\xi_1\,=\,\frac{\Lambda^4}{16\,\pi^2}\,\xi_{01}\,;\quad 
\eta_1\,=\,\frac{\Lambda^4}{16\,\pi^2}\,\eta_{01}\,;\quad
\zeta\,=\,\biggl(\frac{\lambda_V}{\lambda_0}\biggr)^2\,.\nn\\
& &\xi_0\,=\,\frac{\Lambda^4}{16\,\pi^2}\,\xi_{00}\,;\quad 
\eta_0\,=\,\frac{\Lambda^4}{16\,\pi^2}\,\eta_{00}\,.\nn
\eeq
For parameters describing interaction of vector singlet $B$ there is 
no contribution of the triple vertex~(\ref{WWW}), because $B$ is 
an Abelian field. 

Parameters $\xi_i,\,\eta_i$ enter into normalization condition according to 
account of diagram Fig 5. Now this condition takes the form
\be
\frac{N_c\,G^2_1}{16\,\pi^2}\,\log\frac{\Lambda^2}{m^2}\,+\,
\frac{9}{4}\,\Bigl(\xi_1\,+\,2\,\eta_1\Bigr)\,+\,\frac{3}{4}\,
\Bigl(\xi_0\,+\,2\,\eta_0\Bigr)\,=\,1\,;\label{normn}
\ee
From this expression we obtain 
\be
G^2_1\,=\,\frac{(4 - 9 (\xi_1 +2 \eta_1) - 3 (\xi_0 +2 \eta_0))\,4\,\pi^2}
{3\,\Bigl(\log (3\, \pi\,\cos^2\theta_W\,M_H^2)\,-\,
\log (\alpha\,m_t^2)\Bigr)}\,.\label{Gc}
\ee
We take $M_H$~(\ref{hor}) and obtain results, which are 
presented at Table 1. Table 1 contains results of calculations 
for the case of trivial solution of set~(\ref{xi0}), i.e $\xi_0\,=
\,\eta_0\,=\,0$ (all columns but the last); this last column presents 
results for nontrivial solution of 
set~(\ref{xi0}): $\xi_0\,=\,-\,2.781,\,\eta_0\,=\,-\,0.215$.

What are criteria for choosing a solution? 
The main criterion is provided by an energy density of a vacuum. We know, that 
in case of appearance of scalar Higgs condensate $\eta$ the vacuum energy 
density reads
\be
E\,=\,-\,\frac{m_0^4}{4\,\lambda}\,;\label{E}
\ee
where parameters are defined in~(\ref{m0n}), (\ref{lamn}). 
We see, that $\lambda$ is proportional to $G_1^4$, thus the 
smaller is $G_1$ and consequently $m_t$, the deeper becomes 
the energy density minimum. Therefore we have to choose the 
solution, which leads to  the minimal value of $m_t$.

From Table 1 we see, that for calculated value $\lambda_V = - \lambda_0$ 
we have minimal $m_t$ in comparison with two last lines. The last line 
but one corresponds to non-trivial 
solution with $\lambda_V\,=\,0$ and the last line give results for 
completely trivial solution ($\lambda_V\,=\,\xi\,=\,\eta\,=\,0$). Both 
ones give larger values for $m_t$ and thus are not suitable. The 
last column of the table also corresponds to non-minimal values of $G_1$ 
and thus the corresponding solutions are unstable.

However, $m_t \simeq 470\,GeV$ is too large, so let us look for values of 
$\zeta = (\lambda_V/\lambda_0)^2$, which give realistic values of the 
$t$-quark mass. Namely, 
values $\zeta \simeq 1.42 \div 1.43$, 
($\lambda_V\,=\,-\,0.0342$) give  
$G_1$ to be just in correspondence with relation~(\ref{G11}). 
So, contradiction of the normalization condition with value of the 
$t$-quark mass is removed. Thus we come to the conclusion, that we may 
not take into account of the normalization condition in considering 
estimates of composite Higgs mass, as it is done in the previous section. 

How we can interpret the situation, when the desirable value for 
$\lambda_V$ is about 20\% larger, than the calculated one? We would state, 
that corrections to the leading approximation, which is used here, might 
be just of this order of magnitude. To estimate the effect of correction 
let us consider contribution of usual electroweak triple $W$ vertex 
$$
V_{\mu\nu\rho}(p,q,k)\,=\,g\,\Bigl(g_{\mu \nu}(q_\rho -p_\rho)\,+
\,g_{\nu\rho}(k_\mu-q_\mu)\,+\,g_{\rho\mu}(p_\nu-k_\nu)\Bigr);
$$
(where $p,\,q,\,k$ and $\mu,\,\nu,\,\rho$ are as usually incoming momenta 
and indices of $W$s) to compensation equation~(\ref{compW}), which defines 
$\lambda_V$. In addition to diagram Fig. 3 we take three diagrams, in which 
one of vertices is changed to $V_{\mu\nu\rho}$. We have to add also 
contribution of four-$W$ vertex, which is contained in triple vertex
~(\ref{WWW}). This contribution restores the gauge invariance of the result. 
Thus we obtain a contribution, which is proportional to $g/\pi$. 
Namely, instead of~(\ref{compW}) we now have 
$$
\lambda_V\,=\,\lambda_V\,\Biggl(\Biggl(\frac{g \lambda_V\,}{M_W^2}\Biggr)^2\,
\frac{\Lambda^4}{128\,\pi^2}\,+\frac{3\,g}{16\,\pi^2}\,\frac
{g\,\lambda_V\,\Lambda^2}{M_W^2}\Biggr);
$$
That is, if $\lambda_V\,\ne\,0$
\be
1\,=\,\Biggl(\frac{\lambda_V}{\lambda_0}\Biggr)^2\,+\,\frac{3\,g}
{\sqrt{2}\,\pi}\,\frac{\lambda_V}{\lambda_0}\,;\label{compW1}
\ee
and from~(\ref{compW1}) we have
\be
\lambda_V\,\simeq\,\lambda_0\biggl(\pm\,1\,-\,\frac{3\,g}
{2\,\sqrt{2}\,\pi}\biggr)\,.\label{lamcor}
\ee
Value of $g$~(\ref{G11}) gives the correction term here to be about 20\%, 
what corresponds to data presented in Table 1. 
We consider result~(\ref{lamcor}) as qualitative argument for 
possibility of consistent description of data in the framework of our 
approach. Full study of this problem needs 
special efforts. So for the moment one may only state, that value
\be
\lambda_V\,\simeq\,-\,0.034\,;\label{lamV}
\ee
which excellently fits $t$-quark mass, is possible in the approach 
being considered here. 
Sign of $\lambda_V$ is chosen due to experimental limitations 
$\lambda_V\,=\,-\,0.037\,\pm\,0.030$~\cite{Osaka}. We see, that our 
result~(\ref{lamV}) fits this restriction quite nicely. 
We consider this result~(\ref{lamV}) as a prediction for triple gauge 
coupling. 

We have no proof that the possibility being discussed here is the only 
one, which can improve situation with the normalization condition. 
However all other possibilities, which were considered in the course of 
performing the work, lead to wrong sign of corresponding contributions to 
this condition. That is $t$-quark mass becomes even larger, than 
$513\,GeV$. So we are inclined to consider this possibility, especially 
prediction~(\ref{lamV}) as quite promising. In any case we have an 
example of a variant, which has no contradiction with all what we know.

\section{Discussion}

Thus as a result of our study of the model we now come to the theory, 
which describe Standard electroweak interaction of:\\
1. gauge vector bosons sector with input of constants $g$ and 
$\theta_W$;\\
2. heavy quark doublet $t,\,b$ with  the $t$-quark mass 
$174\,\pm\, 5\,GeV$ and the $b$ mass zero;\\
3. necessary composite Higgs sector with mass of surviving neutral 
scalar particle $ M_H\,=\,117\, \pm \,7\,GeV$. 

In addition to these standard interactions there are also effective 
contact interactions, which act in momentum regions limited by 
effective cut-off $\Lambda\,\simeq\,3500\,GeV$.
We calculate the last  number e.g. from relation~(\ref{m0n}). 

Three parameters ($g,\,\theta_W,\,\Lambda$) are the only 
fundamental input of the model. As a technical input we consider 
triple gauge constant $\lambda_V$, which in range of 20\% agrees 
with the calculated value~(\ref{lV}), but its precise value is 
defined by the next approximations. We emphasize, that the presence 
of such interaction with constant~(\ref{lamV}) 
in the framework of the model is necessary.  
We consider result~(\ref{lamV}) as very important prediction of the 
model. Of course, the most important for the model is the prediction, 
that hints for $115\,GeV$ Higgs~\cite{New} have to be confirmed. 

It is worthwhile here to comment the problem of accuracy of the approach. 
Our approximations are the following.\\
1. One-loop diagrams. We estimate accuracy to be around 
$\bar X_1\,\simeq\,0.08$. \\
2. Corrections being of order of magnitude $g/\pi$ give 
uncertainty around 0.2.\\
3. The leading order of $1/N_c$ expansion. Usually precision of 
$1/N_c$ expansion is estimated to be $1/N_c^2\,\simeq\, 0.11$.\\
4. We keep only logarithmic terms. 
Uncertainty is estimated to be $1/\log(\Lambda^2/m_t^2)\,\simeq\,0.16$.

Thus the combined accuracy, provided items 1 -- 4 being independent, 
is estimated to be around 29\%. As a matter of fact, all uncertainties, 
which we have encountered above fit into this range.

For the moment we have only one quark doublet. Remind, that now we 
understand why the $t$-quark is heavy and the $b$-quark is almost 
massless. We consider heavy $t,\,b$ doublet as the most 
fundamental one in the sense, that it defines the main 
part of heavy particles masses. Indeed, composite Higgs scalars are just 
consisting of these quarks. We may immediately introduce all other quarks 
and leptons with zero mass completely in the line   
prescribed by the Standard Model. So at this stage of study of 
the model we deal with massive $W,\,Z,\,t,\,H$ and massless all other 
particles. We may expect,that masses of quarks and leptons will 
appear in subsequent approximations. We have to expect their value to 
be less than 29\% of our heavy masses, that is $m_i\,\leq\,23\,GeV$. All 
the masses known satisfy this restriction. We could expect also, that in 
subsequent approximations Higgs scalars are not composed of only heavy 
quarks, but have admixture of lighter quarks and leptons. 

Additional contact interactions~(\ref{nset}) include 
four-fermion interactions of heavy quarks $t,\,b$. So, there is 
practically no experimental limitations on the corresponding 
coupling constant. Limitations exist only for contact interactions 
of light quarks. Limitations for triple gauge interaction~(\ref{WWW}) 
were mentioned above and were shown to agree with predicted 
value~(\ref{lamV}). New contact interaction~(\ref{ffWW}) 
deals with yet undiscovered Higgs scalars and therefore it 
is not a subject for experimental limitations. As an example 
we estimate cross-section of process $e^+\,e^-\,\to\,H\,H$ 
in Appendix B.

To conclude the author would state, that the variant being studied 
presents a possibility to  formulate realistic electroweak interaction 
without elementary scalars, which are substituted by composite 
effective scalar fields. The dynamics of the variant is defined 
in the framework of the Bogolyubov quasi-averages method. Definite 
predictions make checks of the variant to be quite straightforward.
\bigskip
\begin{flushleft}
{\Large \bf Appendix}
\end{flushleft}
\appendix
\section{Effective cut-off}

Let us consider a theory with combination of four-fermion 
interactions~(\ref{psi}). We take the simplest nontrivial term 
in the Bethe-Salpeter equation for connected four-fermion amplitude, 
corresponding to Lorentz structure 
$\bar \psi_L \gamma_\rho \psi_L\,\bar \psi_R \gamma_\rho \psi_R\,=\,
-\,2\,\bar \psi_L \, \psi_R\,\bar \psi_R \, \psi_L\,$, 
which is presented at Fig. 6. The kernel for this equation 
corresponds to simple loop, which gives the following expression 
\be
K(p,q)\,=\,\imath\,\frac{3\,x\,(y+z)}{16\,\pi^2}\,(p-q)^2\,
 \log\,(p-q)^2\,+\,const\,+\,{\cal O}\,((p-q)^2).\label{kernel}
\ee
We are interested in momentum dependence, so we 
consider just the logarithmic term in the kernel, i.e. the first term 
in Eq.~(\ref{kernel}), which we denote $K_{log}(p,q)$. We shall 
see below, that terms containing $const$ and $const\,(p-q)^2$ 
give zero contribution due to boundary conditions.  
Equation for four-fermion amplitude F(p), corresponding to Fig. 4, 
looks like (after applying of Wick rotation)
\beq
& &F(p)\,=\,-\,\imath \int\,\frac{K_{log}(p,q)\,F(q)}{q^2}\,d^4q\,\,=
\,\beta \int\frac{(p-q)^2\,\log(p-q)^2\,F(q)}{q^2}\,d^4q\,; 
\label{eqf}\\
& &\beta\,=\,\frac{3\,x\,(y+z)}{16\,\pi^2}\,.\nn
\eeq
Equations of such type can be studied in the same way as more 
simple equations with kernels being proportional to $((p-q)^2)^{-1}$. 
The method to solve the latter ones was known a long time 
ago~\cite{AF}. For logarithmic case basic angular integrals are the 
following 
\beq
& &\int d\Omega_4\,\log(p-q)^2\,=\,\pi^2\biggl(\theta(x-y)
\biggl(\frac{y}{x}+2 \log\,x\biggr)\,+\,\theta(y-x)
\biggl(\frac{x}{y}+2 \log\,y\biggr)\biggr)\,;\nn\\
& &\int d\Omega_4\,(p\,q)\,\log(p-q)^2\,=\,\frac{\pi^2}{3}
\biggl(\theta(x-y)
\biggl(\frac{y^2}{x} - 3\,y\biggr)\,+\,\theta(y-x)
\biggl(\frac{x^2}{y} - 3\,x\biggr)\biggr)\,;\label{log}\\
& &x\,=\,p^2,\qquad y\,=\,q^2\,.\nn
\eeq
Integrals~(\ref{log}) are sufficient for angular integrations in 
equation~(\ref{eqf}). So we have
\beq
& &F(x)\,=\,\frac{\beta}{32\,\pi^2}\biggl(\frac{1}{3\,x}\int_0^x
y^2\,F(y)\,dy\,+\,3\int_0^x y\,F(y)\,dy\,+\,2\,\log\,x\,\int_0^x
y\,F(y)\,dy\,+\nn\\
& &+\,2\,x\,\log\,x\,\int_0^x F(y)\,dy\,+\,2\,\int_x^\infty 
y\,\log\,y\,F(y)\,dy\,+\,x \int_x^\infty (3 + 2 \log\,y)\,F(y)\, 
dy\,+\nn\\
& &+\,\frac{x^2}{3}\int_x^\infty \frac{F(y)}{y}\,dy\biggr)\,.\label{eq1}
\eeq
Applying differentiations in proper order, we obtain differential 
equation
\be
\frac{d^3}{dx^3}\,\Bigl(x^2\,\frac{d^3}{dx^3}\,(x\,F(x))\Bigr)\,=
\,-\,\frac{\beta}{8\,\pi^2}\,\frac{F(x)}{x}\,.\label{dif}
\ee
Eq.~(\ref{dif}) is easily transformed to the canonical form of 
Meijer equation~\cite{BE}
\beq
& &\biggl(y\,\frac{d}{dy}+\frac{1}{2}\biggr)\,
\biggl(y\,\frac{d}{dy}\biggr)\,
\biggl(y\,\frac{d}{dy}\biggr)\,
\biggl(y\,\frac{d}{dy}-\frac{1}{2}\biggr)\,
\biggl(y\,\frac{d}{dy}-\frac{1}{2}\biggr)\,
\biggl(y\,\frac{d}{dy}-1\biggr)\,F(y)\,+\nn\\
& &+\,y\,F(y)\,=\,0\,;
\qquad y\,=\,\frac{\beta}{512\,\pi^2}\,x^2\,.\label{Meyer}
\eeq
Equation~(\ref{dif}) (or~(\ref{Meyer})) is equivalent to integral 
equation~(\ref{eq1}) provided proper boundary conditions be 
fulfilled. The first one is connected with convergence of integrals 
at infinity. The most intricate conditions are connected with 
behaviour at $x\to 0$. 
From Eq.~(\ref{Meyer}) we see, that at zero there are the 
following independent asymptotics:
\be
\frac{a_{-1}}{x},\;a_{0\,l}\,\log\,x,\;a_0,\;a_{1\,l}\,x\log\,x,\;
a_1\,x,\;a_2\,x^2\,.\label{x0}
\ee
To obtain conditions for $a_i$ one has to substitute into Eq.~(\ref{eq1})
$$
F(y)\,=\,-\,\frac{8\,\pi^2}{\beta}\,x\,\frac{d^3}{dx^3}\,\Bigl(x^2\,
\frac{d^3}{dx^3}\,(x\,F(x))\Bigr)\,;
$$
and perform integrations by parts. Thus we come to conditions
\be
a_{-1}\,=\,a_{0\,l}\,=\,a_{1\,l}\,=\,0\,.\label{cond}
\ee
These results, in particular, lead to zero values of integrals
$$
\int_0^\infty\,F(y)\,dy\,=\,\int_0^\infty\,y\,F(y)\,dy\,=\,0\,;
$$
that guarantees the absence of contributions of constant terms in the 
initial equation, which we have mentioned above. 

According to properties of Meijer functions~\cite{BE} we have unique solution 
of Eq.~(\ref{Meyer}) with the boundary conditions~(\ref{cond})
\be
F(x)\,=\,\frac{\sqrt{\pi}}{2}\,G_{0\,6}^{3\,0}\Bigl(\,y\,|\,
0,\,\frac{1}{2},\,1\,;\,- \frac{1}{2},\,0,\,\frac{1}{2}\,\Bigr)\,;
\qquad y\,=\,\frac{\beta}{512\,\pi^2}\,x^2\,.
\label{sol}
\ee
This function has all qualities of a form-factor. It is equal to unity 
at the origin and decreases with oscillations at infinity. Effective 
cut-off is  estimated from~(\ref{sol})
\be
\Lambda^2\,=\,\frac{16\,\sqrt{2}\,\pi}{\sqrt{\beta}}\,.
\ee

Thus we demonstrate how the effective cut-off arises. Of course, it 
is done in the quasi-ladder approximation, but we think, that 
an account of non-linearities changes numbers, but does not change 
the situation qualitatively. In the present work we will not go 
further and will not try to connect these considerations with 
realistic numbers.

\section{Pair H production}

It is of interest to estimate possible 
effects of interactions~(\ref{ffWW}, \ref{ffBB}) in future experiments. 
We take as an example 
process $e^+\,e^-\,\to \,H\,H$ connected with these interactions. 
Indeed, due to $W W$, $B B$ loops, interactions~(\ref{ffWW}), (\ref{ffBB})  
lead to vertex 
$\bar e e\,H H$. Simple calculation gives the differential cross-section 
of process $e^+\,e^-\,\to \,H\,H$ due to this vertex
$$
\frac{d \sigma}{d \cos \theta}\,=\,\Biggl(\frac{g^2}{4 \pi}\,\log\,
\frac{\Lambda^2}{s}\Biggr)^2\,\Biggl(\biggl(\xi_1 + \frac{1}{3}\xi_0\,
\tan^2\theta_W\biggr)^2\,+\,\biggl(\frac{4}{3}\,\xi_0\,\tan^2\theta_W
\biggr)^2\Biggr)\,\times
$$
$$
\times\,\frac{\pi\,E\,k^5\,\cos^2\theta\,\sin^2\theta}
{16\,\Lambda^8}\,;
$$
where $E$ is the energy of a beam in c.m.r.f., $s = 4\,E^2$ and 
$ k\,=\,\sqrt{E^2-M_H^2}$. The total cross-section reads
$$
\sigma\,=\,\Biggl(\frac{g^2}{4 \pi}\,\log\,
\frac{\Lambda^2}{s}\Biggr)^2\,\Biggl(\biggl(\xi_1 + \frac{1}{3}\xi_0\,
\tan^2\theta_W\biggr)^2\,+\,\biggl(\frac{4}{3}\,\xi_0\,\tan^2\theta_W
\biggr)^2\Biggr)\,\frac{\pi\,E\,k^5}
{60\,\Lambda^8}\,.
$$
Even for $E\,=\,500\,GeV$ we have for $\xi_1 = 4.54,\,\xi_0 = 0$ 
(see Table 1) 
$\sigma\,\simeq\,5\cdot 10^{-43}\,cm^2$, that is far out of 
experimental possibilities.

\newpage
\begin{center}
{\bf Figure captions}
\end{center}
\bigskip
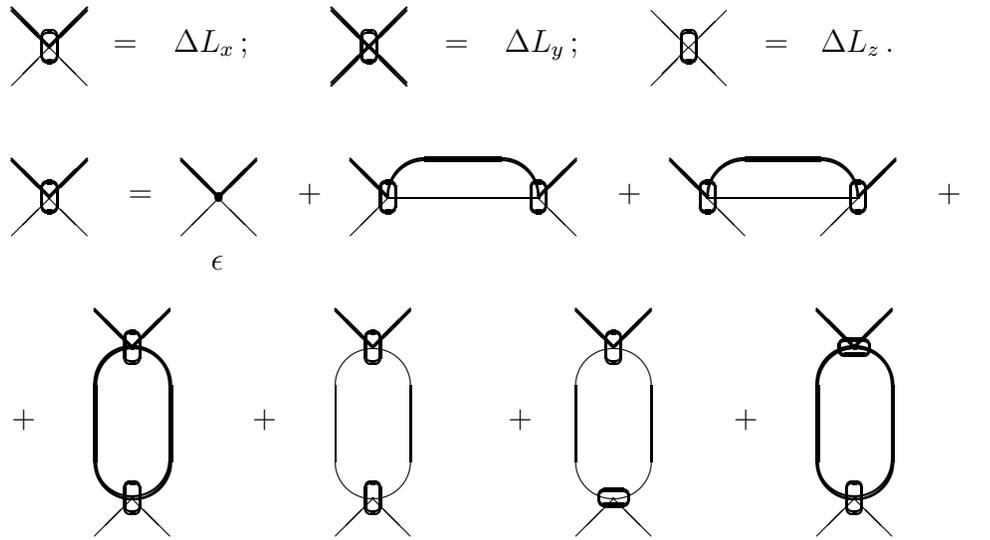
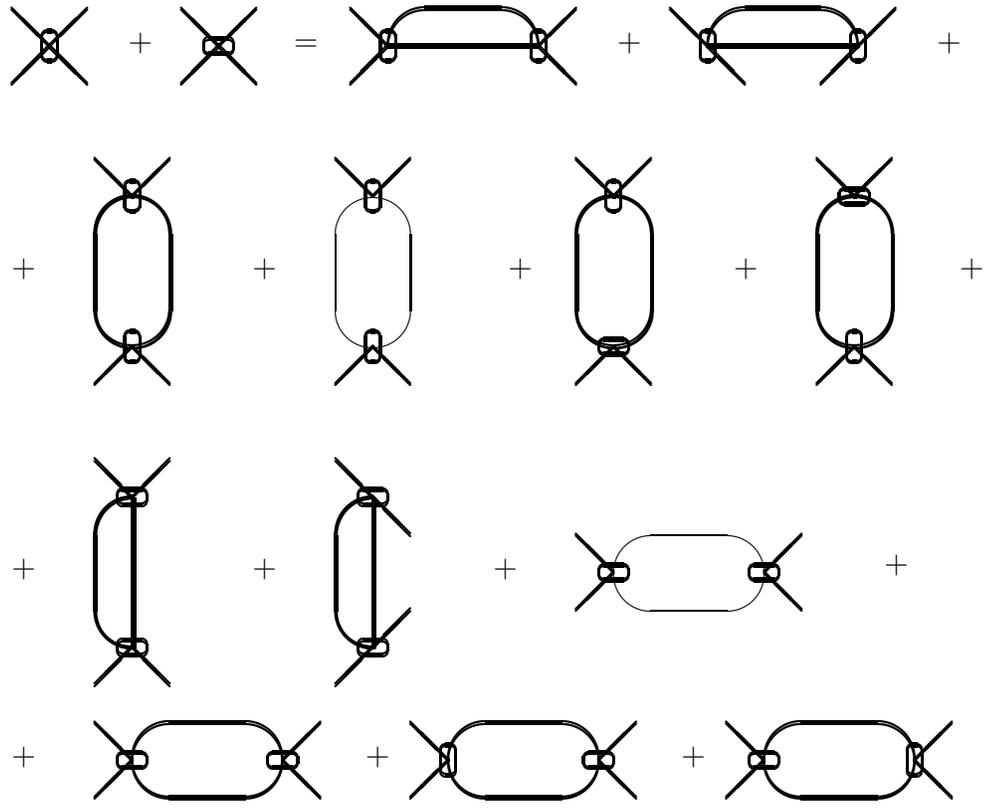
Figure 1. Diagrams for Eq.~(\ref{set}).  
Thick lines represent left spinors $\psi_L$ and thin 
lines represent right ones $\psi_R$; (a) corresponds to the first 
equation of set~(\ref{set}) and (b) corresponds to the second 
equation of~(\ref{set}). The third equation follows from (b) 
by mutual exchange thick $\leftrightarrow$ thin.\\
\\
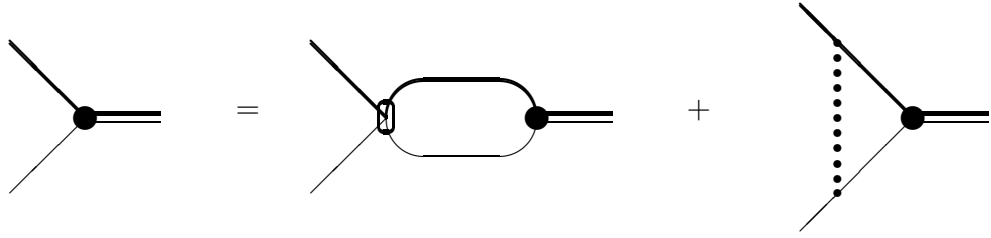
Figure 2. Diagram representation of Bethe-Salpeter equation for scalar 
bound states. Dotted line represents gauge vector field.\\
\\
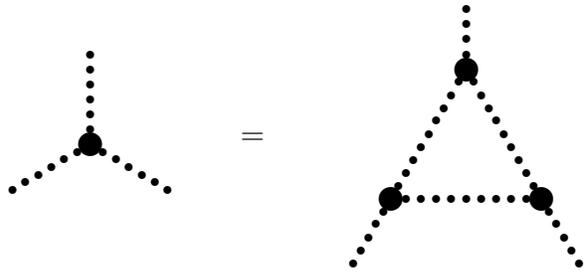
Figure 3. Diagram equation for triple gauge constant $\lambda_V$.\\
\\
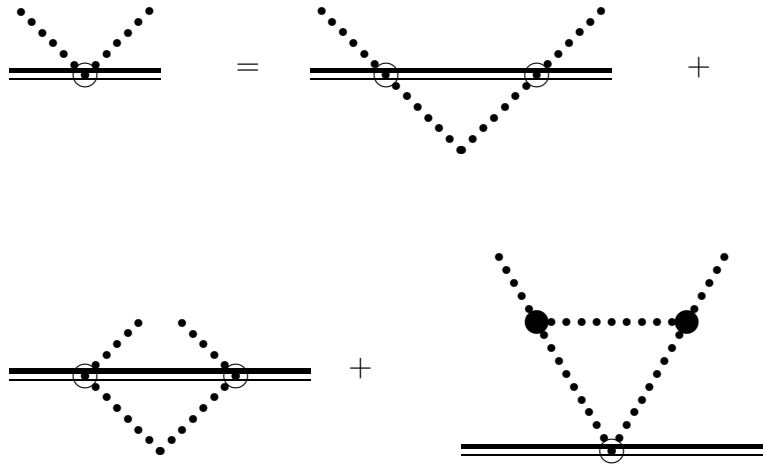
Figure 4. Diagrams representing equations for $\phi^+\phi W W$ vertices.\\
\\
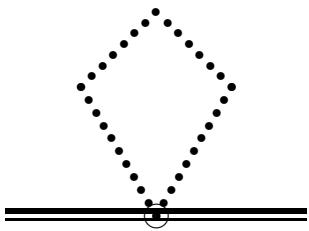
Figure 5. Diagram for contribution of $\phi^+\phi W W$ vertices to the 
normalization condition.\\
\\
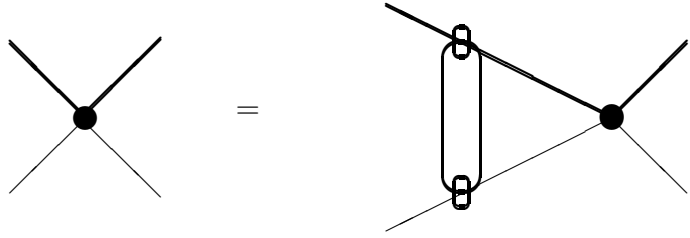
Figure 6. Bethe-Salpeter equation in quasi-ladder approximation.\\
\bigskip
\begin{center}
{\bf Table caption}
\end{center}
\bigskip
Table 1. Yukava coupling $G_1$ and the $t$-quark mass in dependence 
on $\lambda_V$. 
\newpage
\begin{picture}(160,195)
{\thicklines
\put(7.5,192.5){\line(-1,1){5}}
\put(7.5,192.5){\line(1,1){5}}
\put(7.5,192.5){\oval(2,4)}
\put(7.5,192.8){\line(-1,1){5}}
\put(7.5,192.8){\line(1,1){5}}
\put(7.5,192.8){\oval(2,4)}}
\put(7.5,192.5){\line(-1,-1){5}}
\put(7.5,192.5){\line(1,-1){5}}
\put(16,192){=}
\put(24,192){$\Delta L_x\,;$}
{\thicklines
\put(50,192.5){\line(-1,1){5}}
\put(50,192.5){\line(1,1){5}}
\put(50,192.5){\oval(2,4)}
\put(50,192.5){\line(-1,-1){5}}
\put(50,192.5){\line(1,-1){5}}
\put(50,192.8){\line(-1,1){5}}
\put(50,192.8){\line(1,1){5}}
\put(50,192.8){\oval(2,4)}
\put(50,192.8){\line(-1,-1){5}}
\put(50,192.8){\line(1,-1){5}}}
\put(60,192){=}
\put(68,192){$\Delta L_y\,;$}
\put(92.5,192.5){\line(-1,1){5}}
\put(92.5,192.5){\line(1,1){5}}
{\thicklines
\put(92.5,192.5){\oval(2,4)}
\put(92.5,192.8){\oval(2,4)}}
\put(92.5,192.5){\line(-1,-1){5}}
\put(92.5,192.5){\line(1,-1){5}}
\put(102.5,192){=}
\put(110,192){$\Delta L_z\,.$}
{\thicklines
\put(7.5,172.5){\line(-1,1){5}}
\put(7.5,172.5){\line(1,1){5}}
\put(7.5,172.5){\oval(2,4)}
\put(7.5,172.8){\line(-1,1){5}}
\put(7.5,172.8){\line(1,1){5}}
\put(7.5,172.8){\oval(2,4)}}
\put(7.5,172.5){\line(-1,-1){5}}
\put(7.5,172.5){\line(1,-1){5}}
\put(18,172){=}
{\thicklines
\put(30,172.5){\line(-1,1){5}}
\put(30,172.5){\line(1,1){5}}
\put(30,172.5){\circle*{1}}
\put(30,172.8){\line(-1,1){5}}
\put(30,172.8){\line(1,1){5}}
\put(30,172.8){\circle*{1}}}
\put(30,172.5){\line(-1,-1){5}}
\put(30,172.5){\line(1,-1){5}}
\put(29,163){$\epsilon$}
\put(40.5,172){+}
{\thicklines
\put(52.5,172.5){\line(-1,1){5}} 
\put(62.5,172.5){\oval(20,10)[t]}
\put(72.5,172.5){\line(1,1){5}}
\put(52.5,172.5){\oval(2,4)}
\put(72.5,172.5){\oval(2,4)}
\put(52.5,172.8){\line(-1,1){5}} 
\put(62.5,172.8){\oval(20,10)[t]}
\put(72.5,172.8){\line(1,1){5}}
\put(52.5,172.8){\oval(2,4)}
\put(72.5,172.8){\oval(2,4)}}
\put(52.5,172.5){\line(-1,-1){5}} 
\put(72.5,172.5){\line(1,-1){5}}
\put(52.5,172.5){\line(1,0){20}}
\put(83,172){+}
{\thicklines
\put(95,172.5){\line(-1,1){5}} 
\put(105,172.5){\oval(20,10)[t]}
\put(115,172.5){\line(1,1){5}}
\put(95,172.5){\oval(2,4)}
\put(115,172.5){\oval(2,4)}
\put(95,172.8){\line(-1,1){5}} 
\put(105,172.8){\oval(20,10)[t]}
\put(115,172.8){\line(1,1){5}}
\put(95,172.8){\oval(2,4)}
\put(115,172.8){\oval(2,4)}}
\put(95,172.5){\line(1,-1){5}} 
\put(115,172.5){\line(-1,-1){5}}
\put(95,172.5){\line(1,0){20}}
\put(125.5,172){+}
\put(2.5,142){+}
{\thicklines
\put(18.5,152.5){\line(-1,1){5}}
\put(18.5,152.5){\line(1,1){5}}
\put(18.5,142.5){\oval(10,20)}
\put(18.7,142.5){\oval(10,20)}
\put(18.5,152.5){\oval(2,4)}
\put(18.5,132.5){\oval(2,4)}
\put(18.5,152.8){\line(-1,1){5}}
\put(18.5,152.8){\line(1,1){5}}
\put(18.5,142.8){\oval(10,20)}
\put(18.5,152.8){\oval(2,4)}
\put(18.5,132.8){\oval(2,4)}}
\put(18.5,132.5){\line(-1,-1){5}}
\put(18.5,132.5){\line(1,-1){5}}

\put(34.5,142){+}
{\thicklines
\put(50.5,152.5){\line(-1,1){5}}
\put(50.5,152.5){\line(1,1){5}}
\put(50.5,152.5){\oval(2,4)}
\put(50.5,132.5){\oval(2,4)}
\put(50.5,152.8){\line(-1,1){5}}
\put(50.5,152.8){\line(1,1){5}}
\put(50.5,152.8){\oval(2,4)}
\put(50.5,132.8){\oval(2,4)}}
\put(50.5,132.5){\line(-1,-1){5}}
\put(50.5,132.5){\line(1,-1){5}}
\put(50.5,142.5){\oval(10,20)}
\put(68.5,142){+}
{\thicklines
\put(82.5,152.5){\line(-1,1){5}}
\put(82.5,152.5){\line(1,1){5}}
\put(82.5,152.5){\oval(2,4)}
\put(82.5,132.5){\oval(4,2)}
\put(82.5,152.8){\line(-1,1){5}}
\put(82.5,152.8){\line(1,1){5}}
\put(82.5,152.8){\oval(2,4)}
\put(82.5,132.8){\oval(4,2)}}
\put(82.5,132.5){\line(-1,-1){5}}
\put(82.5,132.5){\line(1,-1){5}}
\put(82.5,142.5){\oval(10,20)}
\put(98.5,142){+}
{\thicklines
\put(114.5,152.5){\line(-1,1){5}}
\put(114.5,152.5){\line(1,1){5}}
\put(114.5,142.5){\oval(10,20)}
\put(114.7,142.5){\oval(10,20)}
\put(114.5,152.5){\oval(4,2)}
\put(114.5,132.5){\oval(2,4)}
\put(114.5,152.8){\line(-1,1){5}}
\put(114.5,152.8){\line(1,1){5}}
\put(114.5,142.8){\oval(10,20)}
\put(114.5,152.8){\oval(4,2)}
\put(114.5,132.8){\oval(2,4)}}
\put(114.5,132.5){\line(-1,-1){5}}
\put(114.5,132.5){\line(1,-1){5}}
\put(75,117){(a)}
{\thicklines
\put(7.5,102.5){\line(-1,1){5}}
\put(7.5,102.5){\line(1,1){5}}
\put(7.5,102.5){\oval(2,4)}
\put(7.5,102.5){\line(-1,-1){5}}
\put(7.5,102.5){\line(1,-1){5}}
\put(7.5,102.8){\line(-1,1){5}}
\put(7.5,102.8){\line(1,1){5}}
\put(7.5,102.8){\oval(2,4)}
\put(7.5,102.8){\line(-1,-1){5}}
\put(7.5,102.8){\line(1,-1){5}}}
\put(18,102){+}
{\thicklines
\put(30,102.5){\line(-1,1){5}}
\put(30,102.5){\line(1,1){5}}
\put(30,102.5){\oval(4,2)}
\put(30,102.5){\line(-1,-1){5}}
\put(30,102.5){\line(1,-1){5}}
\put(30,102.8){\line(-1,1){5}}
\put(30,102.8){\line(1,1){5}}
\put(30,102.8){\oval(4,2)}
\put(30,102.8){\line(-1,-1){5}}
\put(30,102.8){\line(1,-1){5}}}
\put(40,102){=}
{\thicklines
\put(52.5,102.5){\line(-1,1){5}} 
\put(62.5,102.5){\oval(20,10)[t]}
\put(72.5,102.5){\line(1,1){5}}
\put(52.5,102.5){\oval(2,4)}
\put(72.5,102.5){\oval(2,4)}
\put(52.5,102.5){\line(-1,-1){5}} 
\put(72.5,102.5){\line(1,-1){5}}
\put(52.5,102.5){\line(1,0){20}}
\put(52.5,102.8){\line(-1,1){5}} 
\put(62.5,102.8){\oval(20,10)[t]}
\put(72.5,102.8){\line(1,1){5}}
\put(52.5,102.8){\oval(2,4)}
\put(72.5,102.8){\oval(2,4)}
\put(52.5,102.8){\line(-1,-1){5}} 
\put(72.5,102.8){\line(1,-1){5}}
\put(52.5,102.8){\line(1,0){20}}}
\put(83,102){+}
{\thicklines
\put(95,102.5){\line(-1,1){5}} 
\put(105,102.5){\oval(20,10)[t]}
\put(115,102.5){\line(1,1){5}}
\put(95,102.5){\oval(2,4)}
\put(115,102.5){\oval(2,4)}
\put(95,102.5){\line(1,-1){5}} 
\put(115,102.5){\line(-1,-1){5}}
\put(95,102.5){\line(1,0){20}}
\put(95,102.8){\line(-1,1){5}} 
\put(105,102.8){\oval(20,10)[t]}
\put(115,102.8){\line(1,1){5}}
\put(95,102.8){\oval(2,4)}
\put(115,102.8){\oval(2,4)}
\put(95,102.8){\line(1,-1){5}} 
\put(115,102.8){\line(-1,-1){5}}
\put(95,102.8){\line(1,0){20}}}
\put(125.5,102){+}
\put(2.5,72){+}
{\thicklines
\put(18.5,82.5){\line(-1,1){5}}
\put(18.5,82.5){\line(1,1){5}}
\put(18.5,72.5){\oval(10,20)}
\put(18.7,72.5){\oval(10,20)}
\put(18.5,62.5){\line(-1,-1){5}}
\put(18.5,62.5){\line(1,-1){5}}
\put(18.5,82.5){\oval(2,4)}
\put(18.5,62.5){\oval(2,4)}
\put(18.5,82.8){\line(-1,1){5}}
\put(18.5,82.8){\line(1,1){5}}
\put(18.5,72.8){\oval(10,20)}
\put(18.5,62.8){\line(-1,-1){5}}
\put(18.5,62.8){\line(1,-1){5}}
\put(18.5,82.8){\oval(2,4)}
\put(18.5,62.8){\oval(2,4)}}
\put(34.5,72){+}
{\thicklines
\put(50.5,82.5){\line(-1,1){5}}
\put(50.5,82.5){\line(1,1){5}}
\put(50.5,62.5){\line(-1,-1){5}}
\put(50.5,62.5){\line(1,-1){5}}
\put(50.5,82.8){\line(-1,1){5}}
\put(50.5,82.8){\line(1,1){5}}
\put(50.5,62.8){\line(-1,-1){5}}
\put(50.5,62.8){\line(1,-1){5}}}
\put(50.5,72.5){\oval(10,20)}
{\thicklines
\put(50.5,82.5){\oval(2,4)}
\put(50.5,62.5){\oval(2,4)}
\put(50.5,82.8){\oval(2,4)}
\put(50.5,62.8){\oval(2,4)}}
\put(68.5,72){+}
{\thicklines
\put(82.5,82.5){\line(-1,1){5}}
\put(82.5,82.5){\line(1,1){5}}
\put(82.5,62.5){\line(-1,-1){5}}
\put(82.5,62.5){\line(1,-1){5}}
\put(82.5,72.5){\oval(10,20)}
\put(82.7,72.5){\oval(10,20)}
\put(82.5,82.5){\oval(2,4)}
\put(82.5,62.5){\oval(4,2)}
\put(82.5,82.8){\line(-1,1){5}}
\put(82.5,82.8){\line(1,1){5}}
\put(82.5,62.8){\line(-1,-1){5}}
\put(82.5,62.8){\line(1,-1){5}}
\put(82.5,72.8){\oval(10,20)}
\put(82.5,82.8){\oval(2,4)}
\put(82.5,62.8){\oval(4,2)}}
\put(98.5,72){+}
{\thicklines
\put(114.5,82.5){\line(-1,1){5}}
\put(114.5,82.5){\line(1,1){5}}
\put(114.5,72.5){\oval(10,20)}
\put(114.7,72.5){\oval(10,20)}
\put(114.5,62.5){\line(-1,-1){5}}
\put(114.5,62.5){\line(1,-1){5}}
\put(114.5,82.5){\oval(4,2)}
\put(114.5,62.5){\oval(2,4)}
\put(114.5,82.8){\line(-1,1){5}}
\put(114.5,82.8){\line(1,1){5}}
\put(114.5,72.8){\oval(10,20)}
\put(114.5,62.8){\line(-1,-1){5}}
\put(114.5,62.8){\line(1,-1){5}}
\put(114.5,82.8){\oval(4,2)}
\put(114.5,62.8){\oval(2,4)}}
\put(128.5,72){+}
\put(2.5,32){+}
{\thicklines
\put(18.5,42.5){\line(-1,1){5}}
\put(18.5,42.5){\line(1,1){5}}
\put(18.5,22.5){\line(-1,-1){5}}
\put(18.5,22.5){\line(1,-1){5}}
\put(18.5,42.5){\line(0,-1){20}}
\put(18.8,42.5){\line(0,-1){20}}
\put(18.5,32.5){\oval(10,20)[l]}
\put(18.7,32.5){\oval(10,20)[l]}
\put(18.5,42.5){\oval(4,2)}
\put(18.5,22.5){\oval(4,2)}
\put(18.5,42.8){\line(-1,1){5}}
\put(18.5,42.8){\line(1,1){5}}
\put(18.5,22.8){\line(-1,-1){5}}
\put(18.5,22.8){\line(1,-1){5}}
\put(18.5,32.8){\oval(10,20)[l]}
\put(18.5,42.8){\oval(4,2)}
\put(18.5,22.8){\oval(4,2)}}
\put(34.5,32){+}
{\thicklines
\put(50.5,42.5){\line(-1,1){5}}
\put(50.5,42.5){\line(1,-1){5}}
\put(50.5,22.5){\line(-1,-1){5}}
\put(50.5,22.5){\line(1,1){5}}
\put(50.5,42.5){\line(0,-1){20}}
\put(50.8,42.5){\line(0,-1){20}}
\put(50.5,32.5){\oval(10,20)[l]}
\put(50.7,32.5){\oval(10,20)[l]}
\put(50.5,42.5){\oval(4,2)}
\put(50.5,22.5){\oval(4,2)}
\put(50.5,42.8){\line(-1,1){5}}
\put(50.5,42.8){\line(1,-1){5}}
\put(50.5,22.8){\line(-1,-1){5}}
\put(50.5,22.8){\line(1,1){5}}
\put(50.5,32.8){\oval(10,20)[l]}
\put(50.5,42.8){\oval(4,2)}
\put(50.5,22.8){\oval(4,2)}}
\put(66.5,32){+}
{\thicklines
\put(82.5,32.5){\line(-1,1){5}}
\put(82.5,32.5){\line(-1,-1){5}}
\put(102.5,32.5){\line(1,1){5}}
\put(102.5,32.5){\line(1,-1){5}}
\put(82.5,32.8){\line(-1,1){5}}
\put(82.5,32.8){\line(-1,-1){5}}
\put(102.5,32.8){\line(1,1){5}}
\put(102.5,32.8){\line(1,-1){5}}}
\put(92.5,32.5){\oval(20,10)}
{\thicklines
\put(82.5,32.5){\oval(4,2)}
\put(102.5,32.5){\oval(4,2)}
\put(82.5,32.8){\oval(4,2)}
\put(102.5,32.8){\oval(4,2)}}
\put(118.5,32.5){+}
\put(2.5,7){+}
{\thicklines
\put(18.5,7.5){\line(-1,1){5}}
\put(18.5,7.5){\line(-1,-1){5}}
\put(38.5,7.5){\line(1,1){5}}
\put(38.5,7.5){\line(1,-1){5}}
\put(28.5,7.5){\oval(20,10)}
\put(18.5,7.5){\oval(4,2)}
\put(38.5,7.5){\oval(4,2)}
\put(18.5,7.8){\line(-1,1){5}}
\put(18.5,7.8){\line(-1,-1){5}}
\put(38.5,7.8){\line(1,1){5}}
\put(38.5,7.8){\line(1,-1){5}}
\put(28.5,7.8){\oval(20,10)}
\put(18.5,7.8){\oval(4,2)}
\put(38.5,7.8){\oval(4,2)}}
\put(49.5,7){+}
{\thicklines
\put(60.5,7.5){\line(-1,1){5}}
\put(60.5,7.5){\line(-1,-1){5}}
\put(80.5,7.5){\line(1,1){5}}
\put(80.5,7.5){\line(1,-1){5}}
\put(70.5,7.5){\oval(20,10)}
\put(60.5,7.5){\oval(2,4)}
\put(80.5,7.5){\oval(4,2)}
\put(60.5,7.8){\line(-1,1){5}}
\put(60.5,7.8){\line(-1,-1){5}}
\put(80.5,7.8){\line(1,1){5}}
\put(80.5,7.8){\line(1,-1){5}}
\put(70.5,7.8){\oval(20,10)}
\put(60.5,7.8){\oval(2,4)}
\put(80.5,7.8){\oval(4,2)}}
\put(91.5,7){+}
{\thicklines
\put(102.5,7.5){\line(-1,1){5}}
\put(102.5,7.5){\line(-1,-1){5}}
\put(122.5,7.5){\line(1,1){5}}
\put(122.5,7.5){\line(1,-1){5}}
\put(112.5,7.5){\oval(20,10)}
\put(102.5,7.5){\oval(4,2)}
\put(122.5,7.5){\oval(2,4)}
\put(102.5,7.8){\line(-1,1){5}}
\put(102.5,7.8){\line(-1,-1){5}}
\put(122.5,7.8){\line(1,1){5}}
\put(122.5,7.8){\line(1,-1){5}}
\put(112.5,7.8){\oval(20,10)}
\put(102.5,7.8){\oval(4,2)}
\put(122.5,7.8){\oval(2,4)}}
\put(72,-3){(b)}
\put(70,-15){Figure 1}
\end{picture}
\newpage
\begin{picture}(160,60)
{\thicklines
\put(10,40){\line(-1,1){10}}
\put(10,40.5){\line(1,0){10}}
\put(10,40.3){\line(-1,1){10}}
\put(10,40.8){\line(1,0){10}}}
\put(10,40){\circle*{3}}
\put(10,40){\line(-1,-1){10}}
\put(10,39.5){\line(1,0){10}}
\put(30,40){=}
{\thicklines
\put(50,40){\line(-1,1){10}}
\put(60,40){\oval(20,10)[t]}
\put(70,40.5){\line(1,0){10}}
\put(50,40){\oval(2,4)}
\put(50,40.3){\line(-1,1){10}}
\put(60,40.3){\oval(20,10)[t]}
\put(70,40.8){\line(1,0){10}}
\put(50,40.3){\oval(2,4)}}
\put(50,40){\line(-1,-1){10}}
\put(60,40){\oval(20,10)[b]}
\put(70,39.5){\line(1,0){10}}
\put(70,40){\circle*{3}}
\put(90,40){+}
{\thicklines
\put(120,40){\line(-1,1){15}}
\put(120,40.5){\line(1,0){10}}
\put(120,40.3){\line(-1,1){15}}
\put(120,40.8){\line(1,0){10}}}
\put(120,40){\line(-1,-1){15}}
\put(120,39.5){\line(1,0){10}}
\put(120,40){\circle*{3}}
\multiput(110,30)(0,2){11}%
{\circle*{1}}
\put(60,10){Figure 2}
\end{picture}
\begin{picture}(160,90)
\multiput(30,40)(0,2){7}%
{\circle*{1}}
\multiput(30,40)(-1.72,-1){7}%
{\circle*{1}}
\multiput(30,40)(1.72,-1){7}%
{\circle*{1}}
\put(30,40){\circle*{3}}
\put(50,40){=}
\multiput(80,50)(0,2){5}%
{\circle*{1}}
\multiput(70,32.8)(2,0){10}%
{\circle*{1}}
\multiput(80,50)(-1,-1.72){16}%
{\circle*{1}}
\multiput(80,50)(1,-1.72){16}%
{\circle*{1}}
\put(80,50){\circle*{3}}
\put(70,32.8){\circle*{3}}
\put(90,32.8){\circle*{3}}
\put(50,10){Figure 3}
\end{picture}
\newpage

\begin{picture}(120,110)
{\thicklines
\put(0,80.5){\line(1,0){20}}
\put(0,80.8){\line(1,0){20}}}
\put(0,79.5){\line(1,0){20}}
\multiput(10,80)(-1.42,1.42){7}%
{\circle*{1}}
\multiput(10,80)(1.42,1.42){7}%
{\circle*{1}}
\put(10,80){\circle{3}}
\put(30,80){=}
{\thicklines
\put(40,80.5){\line(1,0){40}}
\put(40,80.8){\line(1,0){40}}}
\put(40,79.5){\line(1,0){40}}
\multiput(41.42,88.58)(1.42,-1.42){14}%
{\circle*{1}}
\multiput(78.58,88.58)(-1.42,-1.42){14}%
{\circle*{1}}
\put(50,80){\circle{3}}
\put(70,80){\circle{3}}
\put(90,80){+}
{\thicklines
\put(0,40.5){\line(1,0){40}}
\put(0,40.8){\line(1,0){40}}}
\put(0,39.5){\line(1,0){40}}
\multiput(10,40)(1.42,-1.42){8}%
{\circle*{1}}
\multiput(30,40)(-1.42,-1.42){8}%
{\circle*{1}}
\multiput(10,40)(1.42,1.42){6}%
{\circle*{1}}
\multiput(30,40)(-1.42,1.42){6}%
{\circle*{1}}
\put(10,40){\circle{3}}
\put(30,40){\circle{3}}
\put(45,40){+}
{\thicklines
\put(60,30.5){\line(1,0){40}}
\put(60,30.8){\line(1,0){40}}}
\put(60,29.5){\line(1,0){40}}
\put(80,30){\circle{3}}
\multiput(80,30)(-1,1.72){16}%
{\circle*{1}}
\multiput(80,30)(1,1.72){16}%
{\circle*{1}}
\multiput(70,47.2)(2,0){11}%
{\circle*{1}}
\put(70,47.2){\circle*{3}}
\put(90,47.2){\circle*{3}}
\put(45,10){Figure 4}
\end{picture}

\begin{picture}(60,50)
{\thicklines
\put(0,10.5){\line(1,0){40}}
\put(0,10.8){\line(1,0){40}}}
\put(0,9.5){\line(1,0){40}}
\put(20,10){\circle{3}}
\multiput(20,10)(-1,1.72){11}%
{\circle*{1}}
\multiput(20,10)(1,1.72){11}%
{\circle*{1}}
\multiput(10,27.2)(1.42,1.42){8}%
{\circle*{1}}
\multiput(30,27.2)(-1.42,1.42){7}%
{\circle*{1}}
\put(13,-10){Figure 5}
\end{picture}
\newpage
\begin{picture}(160,60)
{\thicklines
\put(10,40){\line(-1,1){10}}
\put(10,40.5){\line(1,1){10}}
\put(10,40.3){\line(-1,1){10}}
\put(10,40.8){\line(1,1){10}}}
\put(10,40){\circle*{3}}
\put(10,40){\line(-1,-1){10}}
\put(10,39.5){\line(1,-1){10}}
\put(30,40){=}
{\thicklines
\put(80,40){\line(-2,1){30}}
\put(60,40){\oval(5,20)}
\put(60,50){\oval(2,4)}
\put(60,30){\oval(2,4)}
\put(80,40){\line(1,1){10}}
\put(80,40){\circle*{3}}
\put(80,40.3){\line(-2,1){30}}
\put(60,40.3){\oval(5,20)}
\put(60,50.3){\oval(2,4)}
\put(60,30.3){\oval(2,4)}
\put(80,40.3){\line(1,1){10}}
\put(80,40.3){\circle*{3}}}
\put(80,40){\line(-2,-1){30}}
\put(80,40){\line(1,-1){10}}
\put(30,10){Figure 6}
\end{picture}
\newpage    
\begin{center}
Table 1.

\bigskip

\begin{tabular}{|l|c|c|l|c|l|c|}\hline
$\zeta$ & $\xi_1$ & $\eta_1$ & $G_1$ & $\lambda_V$ & $m_t\,GeV$& 
$G_1(\xi_0,\eta_0 \ne 0)$\\ \hline
1 & 1.723 & - 0.820 & 2.67 & - 0.0286 &  467.8 & 5.31\\ \hline
1.2 & 3.056 & - 1.421 & 2.13  & - 0.0313 & 373.2 & 5.31 \\ \hline
1.4 & 4.371 & - 1.999 & 1.195 & - 0.0338 & 209.4 & 4.75\\ \hline
1.42 & 4.502 & - 2.056 & 1.047 & - 0.0341 & 183.4 & 4.71 \\ \hline
1.422 & 4.515 & - 2.062 & 1.031 & - 0.0341 & 180.6 & 4.71\\ \hline
1.424 & 4.528 & - 2.068 & 1.015 & - 0.0341 & 177.8 & 4.70\\ \hline
1.426 & 4.541 & - 2.073 & 0.998 & - 0.0342 & 174.8 & 4.70\\ \hline
1.428 & 4.554 & - 2.079 & 0.982 & - 0.0342 & 172.0 & 4.70\\ \hline
1.43 & 4.567 & - 2.085 & 0.964 & - 0.0342 & 168.9 & 4.69\\ \hline
0 & - 2.781 & - 0.215 & 8.392 & 0 & 1470.7 &\\ \hline
0 & 0 & 0 & 2.927 & 0 & 512.8 & \\ \hline
\end{tabular}
\end{center}
\bigskip
\bigskip

\end{document}